\documentstyle[aps,prd]{revtex} 
\begin{document}
\draft
\twocolumn[\hsize\textwidth\columnwidth\hsize\csname
@twocolumnfalse\endcsname
\title{Gravitational Wave Emission from Galactic Radio Pulsars}
\author{J.A. de Freitas Pacheco$^{1,2}$ and J.E. Horvath$^{2}$}
\address{$^1$ Observatoire de la C\^ote D'Azur, B.P. 229 , Nice Cedex 6304, 
FRANCE}

\address{$^2$ Instituto Astron\^omico e Geof\'\i sico, Universidade de S\~ao 
Paulo, Av. M.St\'efano 4200, 04302-904 Agua Funda, S\~ao Paulo SP, BRAZIL}
\date{October,1996}
\maketitle
\begin{abstract}

    We consider in this work 
  continuous gravitational wave (GW) emission from non-axisymmetric radio 
  pulsars. We treat in some detail the observational issues related to the 
  known radio pulsar sample with the aim of unveiling the actual number 
  of sources contributing to GW, which are likely to be the main 
  contributors of GWs. It is shown that the operation of 
  spheroidal GW detectors and full-size interferometers could detect 
  this component of the radiation or impose useful limits on the effective
   oblateness of young radio pulsars.
  
\end{abstract}
\vskip2pc]

\section{Introduction}
  
  Considerable experimental and theoretical efforts are being devoted 
  by several groups to the 
  physics of gravitational waves (hereafter GW). After the celebrated birth 
  of neutrino astronomy [1], 
  GW detection may open 
  a new observational window related to a long-sought prediction of general 
  relativity.
  
  Presently, not only several resonant bars are being operated and continuously 
  improved [2], but also 
  the first full-scale wideband interferometers have 
  been started and should be operational before the turn of the century 
  [3]. The
  expected sensitivity of the advanced versions of 
these detectors in terms of the gravitational
  strain is of the order of $h \approx 10^{-22}$ for short-lived, impulsive
  bursts, and $h \approx 10^{-26}$ for periodic ("long-lived") sources, for
  which integration times of about $10^7$ s are possible. In 
  addition, spheroidal antennas of restricted bandwidth have
  been put forward again to complement the former, specially in 
  the $kHz$ region where they are expected to have their best
  performances [4]. This is precisely a range 
  where we expect to find most of the nearest sources, namely
  rotating neutron stars in the Galaxy. The first 
  positive detection(s) of GW by any of these devices will be
  a profound and significant achievement in physics.
  
 Clearly, a major concern for the above detectors is
  to understand and minimize 
  the sources of noise which define the sensitivity of the antennas at a given 
  frequency. Furthermore, the important question
 of identifying a standard 
 source has been  put forward [5]. Among the possible candidates, 
  it is expected that pulsars, which are known to be abundant in the Galaxy,
  could rank among the most 
  conspicuous emitters of GW. 
Pulsars could have a time-varying 
  quadrupole moment (and hence radiate GW) by either having a slight 
  asymmetry in the equatorial plane (assumed to be 
  orthogonal to the rotation axis) or a 
  misalignment between the symmetry and total angular momentum axes, which 
  produces a wobble in the star motion. In the former case
  the GW frequency is equal to the rotation frequency, whereas in the
  latter two modes are possible: one in which the GWs have the
  same frequency as rotation, and another in which the GWs have twice
  the rotation frequency (the first mode dominates by far at small
  wobble angles while the importance of the second increases for larger
   values). These mechanisms have been considered in the literature
  [6-14] and there seems to be a certain consensus 
  for the likelihood of a positive detection once the antennas
  become operative.

Generally speaking, the calculations and discussions have been 
focused on the detection of {\it individual} sources
  like the Tokyo experiment, devised to detect GW from the Crab
  pulsar [15]. However, very recent work along these 
  lines pointed out that the radiation from the pulsar
  population as a whole could be detectable and paid particular
  attention to the expected time modulation in full-scale
  interferometers [16,17]. On the other hand, omnidirectional
  detectors (which may well be operative on shorter time scales)
  would not experience (by definition) any modulation and will be designed 
  to have a very good sensitivity for continuous radiation and integration
  times of the order of several months. Thus, we have in principle a 
  physical component which is different from simple 
  noise background. 
  We address in this work the expected features of this signal, with particular 
  emphasis on the difficulties created by the use of the observed distribution 
  of the sources, which is affected by several observational biases. The goals 
  of this first approach will be to unveil as much as possible the features 
  of the true population contributing to this emission and to discuss 
  what can be learned from the observations of the latter.
 
\section{Pulsar statistics}
  
  In its simplest form, the evolution of a single pulsar angular
   frequency may be written as 

\begin{equation}  
\omega (t) \, = \, \omega_{o} {\biggl( 1 \, + \, {t \over {\tau_{m}}} 
  \biggr)}^{-m/2} 
\end{equation}

where $\omega_{o}$ is the frequency at birth, $\tau_{m}$ is the
  characteristic e-folding time for the pulsar deceleration
  and $m$ is the index of the torque power-law. From this equation, the
  so-called braking index is defined as
$n \, = \, {\ddot \omega} {\omega} / {\dot \omega}^{2} \, = 1 + 2/m$. 
  As is well known, the case of pure dipolar magnetic braking corresponds
  to $m \, = \, 1$ and hence $ n \, = \, 3$. On the other hand, the
  four presently determined braking indexes differ from the canonical value
  (2.51 for $PSR0531+21$, 2.24 for $PSR0540-69$ and 2.837 for 
  $PSR1509-58$ where the errors affect the last significative digit, see [18] 
and references therein); most 
  notably the recently announced value $1.4 \mp 0.2$ 
for the Vela pulsar by Lyne {\it et al.} [19]. 
  The true reasons for these discrepancies are not clear as yet, but they may 
  reflect substantial non-dipolar  components of the magnetic field {\bf B} 
  [20], time-varying physical parameters like the amplitude
  of {\bf B} [21], the moment of inertia $I$ [19] 
or the angle between the magnetic and rotation
  axes [22,23]. 
  For the moment, we shall leave $m$ unspecified so as
  to allow realistic non-canonical braking scenarios as well. The braking will
  be determined as a byproduct of 
  the analysis of the pulsar period distribution, and therefore should
  be considered in a statistical sense only and not aiming to represent
  any particular object.
  
  To address the expected GW signal from the pulsar population in the
  galactic disk, we should first express the number of 
  {\it observed} pulsars per unit volume and period interval
  $ d^{2}N / dP \, dV$ as 
  a function of pulsar-related parameters and observed quantities. This
  number can be written as

\begin{equation}
 {d^{2}N \over {dP \, dV}} \, = \,{f_B} \, {\nu_{p}(t)} \, {p(R,Z)\over {V_d}}
   {dt \over {dP}}  
\end{equation}
 
    where $\nu_{p}(t)$ is the pulsar birthrate and $V_{d}$ is 
  the disk volume. The factor $f_B$ accounts for the beaming effect
  and for most of the emission models is in the range 0.1 to 0.25.
  The quantity $p(R,Z)$ is the probability to find a pulsar at a galactocentric
  distance $R$ and at a height $Z$ from the galactic plane. The pulsar
  observations at radio frequencies, where they have 
  been widely studied (around $400 \, MHz$), may be used as a "window"
  defining the "observable" volume, since we expect that detection
  decreases considerably for flux densities $S_{400}$ below 10 $mJy$.
  As usual, the flux density (at 400 $MHz$) is defined by
  $S \, = \, L_{p}/ r^{2}$, with $L_{p}$ being the pulsar luminosity and
  $r$ the pulsar distance to the observer. Since 
  the radio sensitivity is necessarily limited by instrumental and signal 
  analysis [24] to those objects emitting stronger than
  $S_{min} \, \simeq \, 10 \, mJy$ (meaning that we have
  an incomplete knowledge of the true sample, see for
  example [25]) the integration over the volume is converted 
  into

  \begin{eqnarray}
 {dN \over {dP}} = {\nu_{p}(t) \over {V_d}} 
  {1 \over {\dot P}}\int^{a}_{-a}2\pi dZ 
\int^{\infty}_{S_{min}} p(Z(S), R(S)){L_{p} \over {2 S^{2}}} dS
\end{eqnarray}  

  To proceed we need to evaluate the integrals in eq.(3). To this 
purpose we have checked 
  that the observed pulsar luminosity function $L_{p}$ can be 
  represented to a very good accuracy by

\begin{equation}
  L_{p} \, = \, A \, {\dot P}^{\alpha} \, exp(-\beta P) \, \, mJy \, 
  kpc^{2} 
\end{equation}

  where $A \, = \, 6.4 \, \times \, 10^{7}$ , $\alpha \, = \, 0.37$ and 
  $\beta \, = \, 0.982 \, s^{-1}$ are the adjusted parameters.
  
  Let us, for the moment, assume an uniform space distribution through
  the disk. Therefore
  using eqs.(2),(3) and (4)
we obtain after integrating

\begin{eqnarray}
  {dN \over {dP}} \, = \, f_{B} \, \nu_{p}(t) \, 
  {\biggl( {A \over {S_{min} \, R_{d}^{2}}} \biggr)} \, 
  {\biggl( {2 \, \tau_{m} \over {m \, P_{o}^{2/m}}} \biggr)}^{(1-\alpha)} \, 
\nonumber\\  
\times {\biggl( {exp(-\beta \, P) \over {P^{(1-2/m)(1-\alpha)}}} \biggr)}
\end{eqnarray}
 
  To extract the value of the unknown parameters, we proceed to plot 
  $log \, dN/dP \, exp(\beta P)$ as a function of $log \, P$ for the actual 
  observed distribution (see, for example, [26,27]). This yields

\begin{equation}
 log \, {\biggl( {dN \over {dP}} exp(\beta P) \biggr)} \, = \, 
  3.00 \, + \, 0.41 \, log \, P  
\end{equation}

  with a correlation coefficient of  0.904 . Therefore, by comparing
eqs.(5) and (6) we get  
  $m \, = \, 1.212$ , and thus a (statistical) braking index $n \, = \, 2.6$, 
  not far from the mean of value
  the directly observed braking indexes [18].  In order to estimate
  the pulsar birth rate [26], we first note that the constant in 
eq.(6) is defined
  by the relation
 
 \begin{equation}
 f_{B}\,\nu_{p}(t)\,{\biggl({A \over{S_{min}\, R_{d}^2}}\biggr)}\,
  {\biggl({2\, \tau_m \over{m\, P_{o}^{2/m}}}\biggr)}^{(1-\alpha)}\, = \, 10^3
\end{equation}  
 
From eq.(6) one obtains

\begin{equation}  
 {{\dot P}\, P^{({2 \over {m}}- 1)}}\, = \, {\biggl({{m\, P_{o}^{2/m}}\over {2\,
   \tau_m}}\biggr)} 
\end{equation}

  Performing  the average of the above equation, using 
  the data from reference [27] yields

\begin{equation}
{\biggl({{m\, P_{o}^{2/m}}\over {2\, \tau_m}}\biggr)} 
  \, = \, 9.9 \times 10^{-15} 
\end{equation}

and inserting this value in eq.(7), one gets 
  $f_B\, \nu_p$\, = \, $10^{- 3}\, yr^{-1}$, which corresponds to a pulsar
  birth rate of once every 100 to 200 yr according to the adopted
  value for the beaming factor. These results are in 
  good agreement with the calculations performed by Lorimer et al. [28].
  On the other hand, assuming an average initial period of pulsars
  of 10 ms, we conclude from eq.(9) that the average
  braking time is  $\tau_m \, \approx$ 1022 $yr$ and that
  their average lifetime is  about 42.3 $Myr$.
  
  Now we have expressed the observed number of pulsars as in eq.(5), 
  we can 
  state that the actual number is simply  $dN/dP$ times $f_{B}^{-1}$, or 
  going back to eq.(2) and changing , for convenience to
  the variable $\omega$

\begin{equation}
  {d^{2}N \over{d \omega \, dV}} \, = \, \lambda \,
  p(Z,R)\, \omega^{-(1+2/m)} 
\end{equation}  
 
  where we have introduced  
  $\lambda \, = \, (2 \, \pi)^{2/m} \, {\nu_{p} \over {V_{d}}} \, 
  {\bigl( {2 \tau_{m} \over {m P_{o}^{2/m}}} \bigr)}$. With 
these results we turn 
  now to the issue of calculating the GW emission expected in the detectors.
  
\section{Expected amplitude of the GW emission}
  
  A population of pulsars having a disk distribution, and for which  
   a slight equatorial asymmetry in each object 
  is expected to be present, produce 
  a non-vanishing value of the gravitational strain
  $ h^{2}$, whose average value is given by 

\begin{equation}  
< h^{2} > \, = \, \int dV \, \int d \Omega \, h_{\ast}^{2} \, 
  {\biggl( {d^{2} N \over {d \omega \, dV}} \biggr)}  
\end{equation}  

  where $h_{\ast}^{2} \, = \, {8 \over {15}} \, 
  {\bigl( {G \over {c^{4}}} \bigr)}^{2} \, {\bar \varepsilon}^{2} \, I_{zz} \, 
  {\Omega^{4} \over {r^{2}}}$ is the angle-averaged contributed amplitude from 
  each source [7,9] 
adequate for omnidirectional antennas, (see [16,17] for a 
  discussion of the interferometer case where the orientation is important), 
  $I_{zz}$ is the principal moment of inertia about the $z$ axis (assumed to 
  be the same for all pulsars) and $\bar \varepsilon \, = \, 
  (a - b)/(ab)^{1/2}$ is a mean over the individual ellipticities in the 
  equatorial plane of the stars ($a,b$ being the radii along the $x$ and $y$ 
  axes respectively).  $\Omega = 2 \pi \nu$ is the GW frequency and is
  equal to twice the pulsar rotation frequency $\omega$ and $r$ is the distance
  to the pulsar. Note that this emission is 
  assumed to arise from non-axisymmetric bodies and {\it not} from precessing 
  ones, which have been addressed in de Ara\'ujo et al.(1994) [14] 
and which are 
  likely to appear as 
impulsive sources (in spite of several remarks along the years 
  some confusion remains in the literature about these cases).
  
  Inserting eq.(10) into eq.(11) and integrating over the volume 
  and bandwidth  $\Delta \Omega$ of a resonant detector we get

\begin{equation}
 < h^{2} >^{1/2} \, = 3.27 \, \times \, 10^{-24} \, \Omega_{c}^{0.675} \, 
  {\Delta \Omega}^{1/2} \, {\bar \varepsilon} 
\end{equation}
 
  where $\Omega_{c}$ is the central frequency of the antenna.
  In order to obtain the numerical factor, we have adopted a disk radius of
  12 kpc and that the nearest pulsar is about 500 pc. Increasing or decreasing
  the latter value by a factor of two produces a decrease or an increase by
  20\% respectively in the coefficient of eq.(12).
  
  The case of broad-band interferometers is a bit more complicated since 
  the signal does depend  on the relative orientation an thus the signal 
  will be modulated (see [16,17]. Although we have not attempted a detailed 
  calculation of this modulation, our model is nevertheless useful to 
  give an estimate of the emission strength {\it modulo} factors of $O(1)$.
   In this case we obtain, after integrating eq.(11) over the range 
  10-1000 $Hz$,

  \begin{equation}
 < h^{2} >^{1/2} \, = \, 3.3 \, \times 10^{-24} \Omega_{max}^{1.175} \, 
  {\bar \varepsilon} 
\end{equation} 

where $\Omega_{max}$ is the upper frequency bound of the antenna. 
  
\section{Discussion}
  
From the results of the former sections we can outline an optimal strategy 
to search for this signal. Let us consider first the case of a pair of 
resonant spheroidal antennae [29]. We assume a design capable of covering 
the frequency range 800-1000 $Hz$ and a strain noise 
$\surd S_{h} \, 10^{-23} \, Hz^{-1/2}$, which seems reasonable for the 
near future. In fact, it would be possible to go as low as 600 $Hz$ for the 
central frequency if the spheroidal detector can be made of niobium 
instead of a cheaper material like $Al$ alloys. Such a reduction is 
important when the whole set of possible sources is considered and is under 
study by some experimental groups [29].

For each of these omnidirectional detectors, the positive identification 
of this GW 
  radiation at, say, $4 \, \sigma$ level requires a minimum detectable 
  amplitude 

\begin{equation} 
h_{c} \, \simeq \, 4 \, {\bigl( {S_{h} \over {\tau}} \bigr)}^{1/2}  
\end{equation}

  where we shall assume that the strain noise $\surd S_{h}$ is 
dominated by thermomechanical forces 
[4,13] 
  and $\tau$ is the integration time $\sim \, 10^{7} \, s$. Imposing 
reasonable technological improvements for a  pair of $Al$ alloy 
" $4^{th}$ "  generation antenna, 
the minimum 
  detectable amplitude is $h_{c} \, \geq \, 10^{-26}$. Comparing 
eq.(12) and eq.(14) 
we deduce that the emission from the ensamble 
  would be detectable by the array if 
  the (mean) characteristic ellipticity of pulsars satisfies

\begin{equation}
{\bar \varepsilon} \, \geq \, 3.6 \, \times \,10^{-7} 
\end{equation}

That is, the array would be sensitive to the pulsar population rotating 
faster than $P \, \simeq \, 2.4 \, ms$. These figure change to 
$\varepsilon \, \geq \, 4.7 \times 10^{-7}$ and $P \, simeq \, 3.6 \, ms$ 
for a niobium array with a central frequency of 600 $Hz$, which gives an 
idea of the range to be explored by these devices. 
  
Since we have always referred to the radio pulsar population, for 
consistency of the picture 
  we shall demand that ${\dot E}_{gw} \, < \, I_{zz} \omega {\dot \omega}$ ,
  where the right term represents the energy rate being extracted from 
  the rotating neutron star.  Using the well-known 
  expression for ${\dot E}_{gw}$ derived by Ferrari and Ruffini [6]

\begin{equation}
{\dot E_{gw}}\, = \, {32 \over 5}\, {G \over {c^5}}\, {I_{zz}^2}\,
  {\varepsilon^2}\,{\omega^{6} } 
\end{equation}

  we get that the GW emission is {\it not} dominant provided that [13] 

\begin{equation}
\varepsilon \, < \, 1.9 \, \times \, 10^{5}\, {({\dot P}{P^3})}^{1/2}
\end{equation} 

  For those target pulsars contributing 
to the pair 
of spheroidal 
antennae we get from eqs.(1) and (8), an 
average initial 
deceleration rate is $\dot P \approx \, 5 \, \times \, 10^{-13}$, and thus
their average equatorial deformation must satisfy
  ${\bar \varepsilon}\, \leq \, 1.5 \, \times \, 10^{-5}$ in order to brake by 
electromagnetic emission. This upper limit is
 a factor of 40 higher then the lower limit implied
  by eq.(15) an)d, in a sense, the window to be probed by the detectors. It 
is interesting to note that the window does not shrink (in fact becomes 
slightly larger) if one considers the lower frequency niobium array.
Although we shall not be concerned 
  with the precise origin of the ellipticity, we note that if the
  crust of neutron stars is solid, then its shape may not necessarily
  be axisymmetric under the effect of rotation, as it would be the
  case for a fluid. More recent studies suggest strongly that most
  of the neutron star is in a liquid phase [30], and in this
  case other mechanisms must be invoked to produce triaxiality.
  Several other mechanisms, however, may induce non-axisymmetric 
  deformations above the detectability bound of eq.(15) as discussed in 
  refs.[31-34] 
(in particular, the presence of stochastic magnetic fields
in the mantle of the neutron star or the existence of a type II superconductor
[35]).

   The possible existence of a subpopulation of 
  silent radio pulsars 
  whose spin-down is driven by GW [13] 
is also interesting and would 
  add to $h$, although we have not based any of our predictions on them. 
We also remark that, given the uncertainties governing our knowledge of 
the young pulsar sample, it is not guaranteed that the initial periods of 
the newborn pulsars are as short as $2.4 \, ms$ (or $3.6 \, ms$ in the 
case of a niobium array). In fact, a direct 
backward extrapolation of the present Crab period gives an initial value of 
$4-5$ times of that, or about $10 \, ms$. If this is the generic case it is 
unlikely that any omnidirectional array could be constructed to detect 
these sources, although interferometers will certainly do due to their 
braod-band coverage and maximum sensitivity at lower frequencies. 
A complicated temporal modulation pattern will arise in the latter, which 
may nevertheless help to extract this signature (see [16]). An 
estimate of the signal strength can be obtained from our model by comparing 
eq.(13) with the maximum sensitivity region of the advanced 
interferometers, giving the 
detectability condition

\begin{equation}
{\bar \varepsilon} \, \geq \, 7 \times 10^{-7}
\end{equation}

to be compared with the upper bound 
${\bar \varepsilon} \, \leq \, 8.5 \times 10^{-5}$. The region to be explored 
is now about two orders-of-magnitude in the ${\bar \varepsilon}$ parameter.
  
To understand what do the detectability conditions really mean we have 
collected in Table 1 the relevant numbers of five "top candidates" according 
to the standard procedure of assuming that $all$ the observed $\dot P$ is due 
to GW emission. This bold assumption yields a maximum ellipticity 
$\varepsilon_{m}$ and therefore the maximum amplitude $h_{m}$ to be expected 
for that individual sources. Table 1 stresses the importance of exploring 
the hidden sources we have been referring to: while the largest $h_{m}$ belong 
to young pulsars like Crab and Vela, the required $\varepsilon_{m}$ seem 
to be rather extreme (a hint for the incorrectness of the "all GW" assumption);
on the other hand $ms$ pulsars like $PSR 1957+20$ can not be strong GW 
emitters because their total energy loses are known to be small. Therefore 
the ideal emitters would be rapidly rotating, high $\dot P$ young pulsars 
which may populate the upper corner of the $\dot P - P$ diagram, having 
the additional advantage of possibly falling in both the spheroidal and 
interferometric detector ranges. It can be said that the 
search for an optimal experimental situation has forced us to address the 
statistical expectations for these yet undiscovered sources.

\begin{table}

\begin{tabular}{lccc}
Source     & $\varepsilon_{m}$  &  $h_{m}$  &  $\nu_{GW}$ [Hz]  \\ 
Vela       & $1.8 \times 10^{-3}$ & $1.9 \times 10^{-24}$ & $22.5$ \\ 
Crab       &  $7.5 \times 10^{-4}$ & $1.5 \times 10^{-24}$ & $60.6$ \\ 
Geminga    & $2.3 \times 10^{-3}$ & $1.2 \times 10^{-24}$ & $8.4$ \\ 
PSR1509-68 & $1.4 \times 10^{-2}$ & $5.8 \times 10^{-25}$ & $13.2$\\ 
PSR1957+20 & $1.6 \times 10^{-9}$ & $1.7 \times 10^{-27}$ & $1244$ \\

\end{tabular}
\vspace{1cm}
\caption{Upper limits to the GW emission from individual 
sources (see text)}
\end{table}

  It is to be noted that, according to their small asymmetry as deduced from 
their observed features [13], the 
millisecond pulsar subpopulation would not contribute to the ensamble 
radiation. Consistently, this subpopulation has {\it not} 
been included when considering the statistical features exposed in Sec.2. 
This does not mean, however, that $ms$ pulsars are not interesting for 
GW searches (see, for example, [13,14]), 
but rather that our 
calculations do not depend on them. We believe that the accumulated evidence 
for a distinct evolutionary path for these objects is enough for a 
separate treatment of their emission.
  
  As a final remark it is important to stress that, in spite of the statistical 
  treatment given in this work, $< h^{2} >^{1/2}$ will be dominated by the 
  youngest pulsars in the Galaxy because of the rather strong
  dependence of the gravitational emission on $\omega$. For example, 
  from our statistical analysis we found that the total number of
  pulsars with $P \leq 4 \, s$  is around $10^{5}$, but only $\sim \, 25$
  objects would be responsible for 70\% of the total gravitational
  radiation in the interferometers. We would also predict in our model 
that only $\sim \, 3$ of them would be
  detected at radio frequencies (400 $Mhz$) in the period range 10-30 ms.
The true situation may be more optimistic than we think. According to 
Camilo [36], a clue about the young pulsar population may be hidden in the 
$\dot P - P$ diagram. His argument is precisely that evolution along 
a {\bf B} $= \, constant$ line may require the presence of many young 
pulsars with $\dot P \, \geq \, 10^{-14} s/s$ and $P \, \leq \, 50 \, ms$ 
which remain undiscovered because of selection effects and are missed in 
the existing searches. This independent evidence 
reinforces the motivation of our statistical approaches and calls for future 
refinements.
  Detection of this GW radiation would be an important check for
   the first GW observatories, and even a non-detection would be
   useful to set constraints on the shape an on the
  statistics of  young  galactic pulsars.

   J.E.H. wishes to acknowledge the hospitality and 
  financial support from Department Fresnel, O.C.A. during a visit which 
  made this work possible and a Research Fellowship from the CNPq Agency 
 (Brazil).

\end{document}